# ExoTiC-ISM: A Python package for marginalised exoplanet transit parameters across a grid of systematic instrument models

**Iva Laginja**[1, 2, 3] **and Hannah R. Wakeford**[4]

**1** Space Telescope Science Institute, Baltimore, USA **2** DOTA, ONERA, Université Paris Saclay, F-92322 Châtillon, France **3** Aix Marseille Université, CNRS, CNES, LAM, Marseille, France **4** School of Physics, University of Bristol, HH Wills Physics Laboratory, Tyndall Avenue, Bristol BS8 1TL, UK

## Science background

### Transit spectroscopy of exoplanets

There have been a slew of planet detections outside our own solar system over the past two decades and several characterisation methods can be used to determine their chemical compositions. One of them is transit spectroscopy. With this technique, astronomers measure the star light passing through an exoplanet's atmosphere while it is transiting in front of its host star. Imprinted on this light are the absorption signatures of different materials - atoms and molecules in the gas phase, or solid or liquid aerosols - in the transiting planet's atmosphere. Using a spectrograph the flux is recorded as a function of wavelength, allowing scientists to construct absorption/transmission spectra, with the goal of identifying the chemical composition of the atmosphere.

There are many different chemical components that transit spectroscopy can reveal in an exoplanet, but a majority of the exoplanets studied via transmission spectroscopy are on close-in orbits around their stars lasting only several days, most of them giant Jupiter- or Neptune-sized worlds. For these giant, close-in exoplanets, the most dominant source of absorption will be from water vapour, which is expected to be well-mixed throughout their atmosphere. $H_2O$ has strong absorption in the near-infrared (IR) with broad peaks at 0.9, 1.4, 1.9, and 2.7 $\mu$m. However, these absorption features cannot be measured from the ground as the Earth's atmosphere, filled with water vapour, gets in the way. To measure $H_2O$ in the atmospheres of exoplanets, astronomers use the Hubble Space Telescope's Wide Field Camera 3 (HST WFC3) infrared capabilities to detect the absorption signatures of $H_2O$ at 0.9 $\mu$m with the G102 grism, and at 1.4 $\mu$m with the G141 grism (e.g. Kreidberg et al., 2015; Sing et al., 2016; Spake et al., 2018; Wakeford et al., 2018, 2017).

### Calibration of instrument systematics with marginalisation

While all telescopes aim at recording the light of distant worlds as accurately as possible, every instrument will have its own signature that it superimposes on the collected signal. These effects that contaminate our data are called "instrument systematics" and need to be calibrated out before an observation can be interpreted truthfully. All the different systematics that influence the result recorded by an instrument are combined into a systematic instrument model to be used during the calibration. These models are not always obvious and different authors often suggest differing systematic instrument models being applied to data from the same instrument, often favouring models that work particularly well for any given data set. This makes it hard to compare data sets to each other, yielding moderately different results for the parameters of a transiting planet when a different systematic model was applied to the data.

The solution that was applied to WFC3 data in Wakeford, Sing, Evans, Deming, & Mandell (2016) performs a marginalisation across a grid of systematic models that take different corrections across an exoplanet transit data set into account. Following the method proposed by Gibson (2014), a Levenberg-Marquardt least-squares minimisation is performed across all systematic models, which yields a set of fitted transit parameters for each systematic model. We then use the resulting Akaike Information Criterion (AIC) to calculate each model's evidence (marginal likelihood) and normalised weight. These weights are then used to calculate the marginalised fit parameters, leading to results that will not depend as heavily on the individual choice of systematic model. Finally, performing this for each lightcurve constructed at each wavelength from the measured spectrum results in the measured transmission spectrum of the exoplanet.

## The `ExoTiC-ISM` package

### Functionality

`ExoTiC-ISM` (Exoplanet Timeseries Characterisation - Instrument Systematic Marginalisation) is an open-source Python package that computes the transit depth from a timeseries lightcurve, while sampling a grid of pseudo-stochastic models to account for instrument based systematics that may impact the measurement, following the method proposed by Gibson (2014) and implemented by Wakeford et al. (2016). While there are a number of Python solutions to create and fit transiting planet light curves, `ExoTiC-ISM` is a lightcurve fitting tool that focuses particularly on the statistical method of marginalisation. It allows for a transparent method of data analysis of systematics impacting a measurement. While other methods, such as Gaussian processes (GP), can account for the likelyhood of systematics impacting your measurement, these methods can typically not easily determine which systematics are the most important, and which combination of systematics is specifically affecting your data set. `ExoTiC-ISM` allows you to evaluate a grid of instrument systematic models to obtain the needed information on the dominant systematics enabling you to design the next observation to be more efficient and precise. As the authors of the original method paper state (Wakeford et al., 2016): "The use of marginalisation allows for transparent interpretation and understanding of the instrument and the impact of each systematic [model] evaluated statistically for each data set, expanding the ability to make true and comprehensive comparisons between exoplanet atmospheres."

The currently implemented instrument systematic grid is composed of a series of 49 polynomial functions that are specifically designed to account for systematics associated with the detectors on HST WFC3 (Wakeford et al., 2016). However, this can be adapted to other instruments. The package performs the Levenberg-Marquardt least-squares minimisation across all models with the `sherpa` package (Burke et al., 2019; Doe et al., 2007; Freeman, Doe, & Siemiginowska, 2001) for modeling and fitting data, and then calculates the AIC and normalised weight to marginalise over the fit parameters (e.g. transit depth $rl$, inclination $i$, a/$R_*$, center of transit time) using each systematic model. This method is different from evaluating each systematic model independently and selecting the "best" one purely by minimising the scatter of its residuals as that would not include a penalisation for increased model complexity nor information from similarly likely systematic corrections.

The original code was written in IDL, which was used to publish marginalised transit parameters for five different exoplanets (Wakeford et al., 2016) observed in the IR with the G141 grism on HST's WFC3. The `ExoTiC-ISM` package described in this paper implements a marginalisation for that same grism and extends its functionality to the G102 grism, which uses the same grid of systematic models (see results by Wakeford et al., 2017, 2018).

## Dependencies and usage

`ExoTiC-ISM` is written in Python with support for Python 3.6 and 3.7 on MacOS and Linux. It makes use of the packages `numpy` (Oliphant, 2006; van der Walt, Colbert, & Varoquaux, 2011), `astropy` (Astropy Collaboration et al., 2013; Price-Whelan et al., 2018), `pandas` (McKinney, 2010; Reback et al., 2019), `matplotlib` (Caswell et al., 2019; Hunter, 2007), `sherpa` (Burke et al., 2019; Doe et al., 2007; Freeman et al., 2001), as well as some custom functions, like an implementation of the transit function by Mandel & Agol (2002). It applies a 4-parameter limb darkening law as outlined in Claret (2000) and Sing (2010), using either the 1D Kurucz stellar models or the 3D stellar atmosphere models by Magic, Chiavassa, Collet, & Asplund (2015).

The required inputs for the analysis are two text files that contain the uncorrected lightcurve of the observed object, and a wavelength array. Input parameters are rendered from a `config.ini` file and we provide an `environment.yml` file to build a conda environment to run the package in. Development in Python and hosting the repository on GitHub will facilitate the usage of the package by researchers, as well as further functional development; an introductory tutorial is provided in the form of a Jupyter Notebook.

## Outlook

While its current capabilities are limited to WFC3 data taken with the G141 and G102 grism, the package's functionality will be extended to the UVIS G280 grism (Wakeford et al., 2020) and the G430L and G750L gratings of the Space Telescope Imaging Spectrograph (STIS) on HST. This will lay the groundwork for the envisioned future extension to implement systematic grids for select instruments on the James Webb Space Telescope (JWST) and obtain robust transit spectra for JWST data.

## Acknowledgements

The authors would like to acknowledge Matthew Hill who translated the first part of the IDL version of this code to Python. We also acknowledge contributions by Tom J. Wilson to the statistical testing of the code. We also thank the Sherpa team for their fast and detailed responses to questions we had during the implementation of their package. This work is based on observations made with the NASA/ESA Hubble Space Telescope, HST-GO-14918, that were obtained at the Space Telescope Science Institute, which is operated by the Association of Universities for Research in Astronomy, Inc.